\documentclass[preprint,preprintnumbers, prd, floatfix, superscriptaddress,nofootinbib] {revtex4-1}
\usepackage{epsfig}
\usepackage{subfigure}
\usepackage{dcolumn}
\usepackage{bm}
\usepackage[usenames ,dvipsnames]{xcolor}
\usepackage{slashed}
\usepackage{graphicx,color}
\usepackage{hyperref}

\begin{document}

\title{Testing the $W$-exchange mechanism\\with two-body baryonic $B$ decays}

\author{Y.K. Hsiao}
\email{yukuohsiao@gmail.com}
\affiliation{School of Physics and Information Engineering, 
Shanxi Normal University, Linfen 041004, China}

\author{Shang-Yuu Tsai}
\email{shangyuu@gmail.com}
\affiliation{School of Physics and Information Engineering, 
Shanxi Normal University, Linfen 041004, China}

\author{Chong-Chung Lih}
\email{cclih@phys.nthu.edu.tw}
\affiliation{Department of Optometry, Central Taiwan University of Science and Technology, 
Taichung 40601, Taiwan}

\author{Eduardo Rodrigues}
\email{eduardo.rodrigues@liverpool.ac.uk}
\affiliation{Oliver Lodge Laboratory, University of Liverpool, Liverpool, U.K.}


\begin{abstract}
The role of $W$-exchange diagrams in baryonic $B$ decays is poorly understood,
and often taken as insignificant and neglected.
We show that charmful two-body baryonic $B\to {\bf B}_c \bar {\bf B}'$ decays
provide a good test-bed for the study of the $W$-exchange topology,
whose contribution is found to be non-negligible;
here ${\bf B}_c$ is an anti-triplet or a sextet charmed baryon,
and ${\bf \bar B}'$ an octet charmless (anti-)baryon.
In particular, we calculate that
${\cal B}(\bar B^0\to\Sigma_c^{+}\bar p)=(2.9^{+0.8}_{-0.9})\times 10^{-6}$
in good agreement with the experimental upper bound.
Its cousin $\bar B_s^0$ mode, $\bar B_s^0\to\Lambda^+_c\bar p$, is a purely
$W$-exchange decay, hence is naturally suited for the study of the role of
the $W$-exchange topology. We predict
${\cal B}(\bar B_s^0\to\Lambda^+_c\bar p)=(0.8\pm 0.3)\times 10^{-6}$,
a relatively large branching ratio to be tested with a future measurement
by the LHCb collaboration.
Other predictions, such as
${\cal B}(\bar B^0\to\Xi_c^+\bar\Sigma^-)=(1.1\pm 0.4)\times 10^{-5}$,
can be tested with future Belle II measurements.
\end{abstract}

\maketitle

\section{Introduction}
Decays of $B$ mesons to multi-body baryonic final states, such as
$B\to{\bf B\bar B'}M$ and ${\bf B\bar B'}MM'$,
where ${\bf B}$ ($M$) represents a baryon (meson), have been richly studied.
Their branching ratios are typically at the level of $10^{-6}$~\cite{pdg}.
These relatively large branching ratios are due to the fact that
the baryon-pair production tends to occur in the threshold region
of $m_{\bf B\bar B'}\simeq m_{\bf B}+m_{\bf \bar B'}$, where
the threshold effect with a sharply raising peak
can enhance the branching ratio~\cite{Wei:2007fg,Aaij:2017vnw,Aaij:2017pgn,Lu:2018qbw}.
On the other hand, without the recoiled meson(s) to carry away the large energy release,
the two-body $B\to{\bf B\bar B'}$ decays proceed at the $m_B$ scale,
several GeV away from the threshold region,
resulting in the suppression of their decay rates~\cite{Hou:2000bz,Suzuki:2006nn}.
So far, only three charmless modes have been seen experimentally:
$\bar B^0\to p\bar p$ (${\cal B} \simeq {\cal O}(10^{-8})$) and
$B^-\to \Lambda^{(*)}\bar p$ (${\cal B} \simeq {\cal O}(10^{-7})$),
with $\Lambda^*\equiv \Lambda(1520)$~\cite{Aaij:2017gum,Aaij:2016xfa,Aaij:2013fla}.

\begin{figure}[t!]
\centering
\includegraphics[width=3.3in]{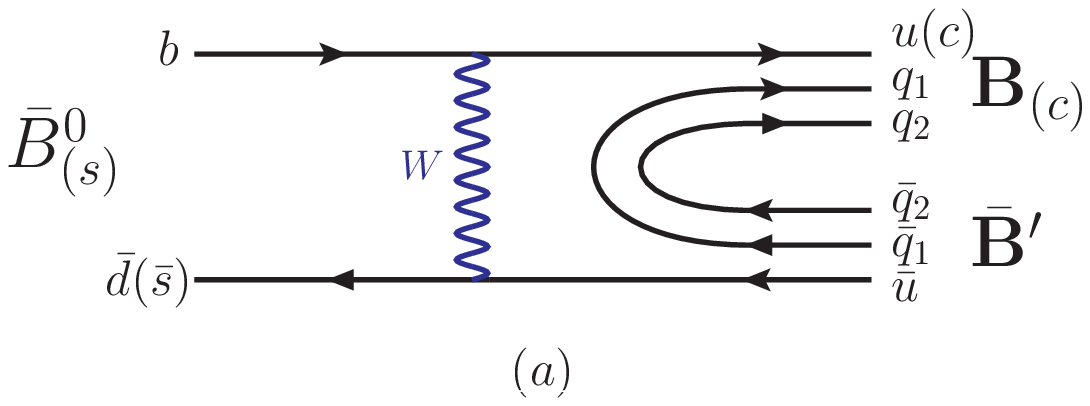}
\includegraphics[width=3.3in]{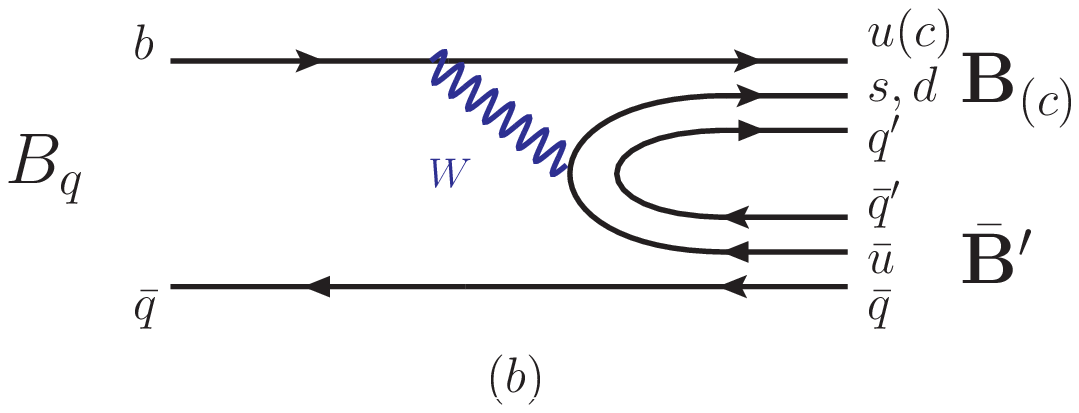}
\includegraphics[width=3.3in]{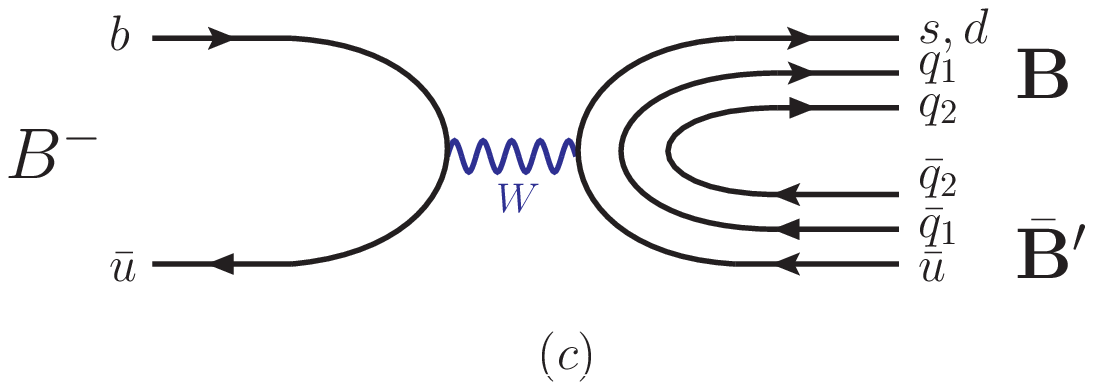}
\caption{Feynman diagrams for the $B\to{\bf B}_{(c)} \bar {\bf B}'$ decays,
where (a,c) depict the $W$-exchange and annihilation processes,
respectively, and (b) depicts the $W$-emission process,
with $q=(u,d,s)$ for $B_q\equiv(\bar B^0,B^-,\bar B^0_s)$.}\label{fig:Feynman}
\end{figure}

The tree-level dominated $B\to {\bf B\bar B'}$ decays
can proceed through
the $W$-exchange, emission and annihilation diagrams, depicted
in Figs.~\ref{fig:Feynman}(a,b,c), respectively.
However, the $W$-exchange (annihilation) process is
regarded as helicity suppressed~\cite{Chen:2008pf,Bevan:2014iga},
and hence neglected in theoretical studies~\cite{Chernyak:1990ag,Ball:1990fw,
Chang:2001jt,Cheng:2001tr,Chua:2013zga,He:2006vz}.
Moreover, one also neglects
the penguin-level gluon-exchange (annihilation) contributions,
which leads to ${\cal B}(\bar B_s^0\to p\bar p)\simeq 0$~\cite{Chua:2013zga}.
Consequently, the observation of the $\bar B_s^0\to p\bar p$ decay would provide
valuable information on whether contributions from the
exchange (annihilation) processes play a significant role.
The smallness of the current upper bound on its branching ratio,
$\mathcal{B}(\bar B_s^0\to p\bar p)<1.5\times 10^{-8}$~\cite{Aaij:2017gum},
indicates an experimentally difficult decay mode to study
the role of exchange (annihilation) diagrams.

%

On the other hand, experimental data show that
${\cal B}(B\to {\bf B}_c {\bf\bar B}')\simeq (10^2-10^3){\cal B}(B\to{\bf B\bar B'})$,
where ${\bf B}_c$ denotes a charmed baryon~\cite{pdg}.
With significantly larger decay rates, charmful two-body baryonic $B$ decays
offer an interesting and suitable environment in which to study and test
the role of the $W$-exchange (annihilation) mechanism.
The set of measured $B\to {\bf B}_c {\bf\bar B}'$ branching ratios is
nevertheless scarce~\cite{pdg,Gabyshev:2002dt,Aubert:2008ax}:
\begin{eqnarray}\label{BcBdata}
&&{\cal B}(\bar B^0\to \Lambda_c^+ \bar p)=(1.54\pm 0.18)\times 10^{-5}\,,\nonumber\\
&&{\cal B}(\bar B^0\to \Sigma_c^+ \bar p)<2.4\times 10^{-5}\,,\nonumber\\
&&{\cal B}(B^-\to \Sigma_c^0 \bar p)=
(2.9\pm 0.7)\times 10^{-5}\,.
\end{eqnarray}
As in the case of the charmless final states considered above,
the $B\to {\bf B}_c {\bf\bar B}'$ decays
can proceed through both the $W$-exchange and $W$-emission diagrams.
%
Again, theoretical studies regard the $W$-exchange diagram as helicity-suppressed,
which is in analogy with the leptonic $B\to\ell\bar \nu_\ell$ decays,
and take the $W$-emission diagram
as the dominant contribution~\cite{Chernyak:1990ag,Ball:1990fw,
Cheng:2001ub,Cheng:2002sa,He:2006vz}.
For clarification, we present
the amplitudes of $B\to\ell\bar \nu_\ell$ and $B\to{\bf B}_c\bar {\bf B}'$
with $W$-exchange contribution
in Fig.~\ref{fig:Feynman}a as
\begin{eqnarray}\label{helicity}
{\cal A}(B\to\ell\bar \nu_\ell)&\propto& m_\ell \bar u(1+\gamma_5)v\,,\nonumber\\
{\cal A}(B\to{\bf B}_c\bar {\bf B}')&\propto&
m_c \langle {\bf B}_c\bar {\bf B}'|\bar c(1+\gamma_5)q|0\rangle\,,
\end{eqnarray}
where ${\cal A}(B\to{\bf B}_c\bar {\bf B}')$ by equation of motion
is presented as a reduced form with the quark mass $m_c$~\cite{Hsiao:2014zza}.
In Eq.~(\ref{helicity}),
the small $m_\ell$, with $\ell=(e,\mu)$, is responsible
for the helicity suppression in $B\to\ell\bar \nu_\ell$.
Nonetheless,
$m_c\sim1.3$~GeV~\cite{pdg} in ${\cal A}(B\to{\bf B}_c\bar {\bf B}')$
is clearly helicity allowed,
indicating that neglecting its contribution may not be a valid assumption to make.
For completeness, we note that the theoretical studies
in Refs.~\cite{Bediaga:1991eu,Pham:1980dc,Pham:1980xe,Hsiao:2014zza}
also considered the exchange and annihilation contributions
in $B\to{\bf B\bar B'}$ and $D_s^+\to p\bar n$.

We therefore propose the study of the family of charmful two-body baryonic
$B\to{\bf B}_c\bar {\bf B}'$ decays to improve our knowledge of the
role of $W$-exchange diagrams in $B$ decays to baryonic final states.
Since modes such as $\bar B^0\to\Xi_c^+\bar\Sigma^-$ and $\bar B_s^0\to\Lambda^+_c\bar p$
can only proceed via the $W$-exchange diagram,
measurements of their branching ratios are direct tests of the $W$-exchange mechanism.

\section{Formalism}
Besides the $W$-emission~(${\cal A}_{\rm em}$) amplitudes
studied elsewhere~\cite{Cheng:2001ub,Cheng:2002sa,He:2006vz},
we consider the often neglected $W$-exchange~(${\cal A}_{\rm ex}$)
amplitudes for the charmful two-body baryonic $\bar B^0_{(s)}\to {\bf B}_c{\bf \bar B}'$ decays,
where ${\bf B}_{c}$ denotes the anti-triplet and the sextet charmed baryon states,
$(\Xi_c^{+,0},\Lambda_c^+)$ and $(\Sigma_c^{++,+,0},\Xi_c^{\prime +,0},\Omega_c^0)$, respectively,
and ${\bf \bar B}'$ an octet charmless (anti-)baryon.
The decays with the decuplet charmless (anti-)baryons are excluded
from the calculations in this paper
due to the lack of the corresponding timelike baryon form factors.

We show in Table~\ref{tab:classification} the amplitudes involved in the interested
$\bar B^0_{(s)}\to {\bf B}_c\bar {\bf B}'$ modes.
The decay rate of modes that can only occur through the $W$-exchange diagram
would be vanishingly small by construction if the importance of these diagrams
was to be insignificant:
\begin{eqnarray}
&&{\cal B}(\bar B^0\to \Xi_c^+\bar\Sigma^-,\Sigma_c^{++}\bar\Delta^{--},
\Xi_c^{\prime +}\bar\Sigma^-,\Omega_c^0\bar\Xi^0)=0\,,\nonumber\\
&&{\cal B}(\bar B^0_s\to \Lambda_c^+\bar p,\,\Lambda_c^+\bar\Delta^-,
\Sigma_c^{++}\bar\Delta^{--},
\Sigma_c^+\bar p,\,\Sigma_c^+\bar\Delta^-,
\Sigma_c^0\bar n,\,\Sigma_c^0\bar\Delta^0)=0\,.
\end{eqnarray}
None of these relations has yet been verified experimentally.

\begin{table}[t!]
\caption{Classification for $B\to {\bf B}_c\bar {\bf B}'$ decays, where
${\cal A}_{\rm ex}$ and ${\cal A}_{\rm em}$ denote the amplitudes through
the $W$-exchange and $W$-emission diagrams, respectively.}
\label{tab:classification}
\begin{tabular}{llll}
\hline
$\bar B^0\to{\bf B}_c\bar{\bf B}'$ & Amplitudes & $\bar B^0\to{\bf B}_c\bar{\bf B}'$ & Amplitudes \\
\hline
$\Lambda_c^+\bar p,\,\Lambda_c^+\bar\Delta^-$ & $\mathcal{A}_{\rm ex}+\mathcal{A}_{\rm em}$ &
$\Sigma_c^{++}\bar\Delta^{--}$ & $\mathcal{A}_{\rm ex}$ \\
$\Xi_c^+\bar\Sigma^-$ & $\mathcal{A}_{\rm ex}$ & $\Sigma_c^+\bar p,\,\Sigma_c^+\bar\Delta^-$ &
$\mathcal{A}_{\rm ex}+\mathcal{A}_{\rm em}$ \\
$\Xi_c^0\bar\Lambda,\,\Xi_c^0\bar\Sigma^0$ & $\mathcal{A}_{\rm ex}+\mathcal{A}_{\rm em}$  &
$\Sigma_c^0\bar n,\,\Sigma_c^0\bar\Delta^0$ &  $\mathcal{A}_{\rm ex}+\mathcal{A}_{\rm em}$ \\
&  & $\Xi_c^{'+}\bar\Sigma^-$ &  $\mathcal{A}_{\rm ex}$ \\
&   & $\Xi_c^{'0}\bar\Lambda,\,\Xi_c^{'0}\bar\Sigma^0$  &
$\mathcal{A}_{\rm ex}+\mathcal{A}_{\rm em}$ \\
&   &  $\Omega_c^0\bar\Xi^0$ & $\mathcal{A}_{\rm ex}$  \\
\hline
\hline
$\bar B_s^0\to\mathbf{B}_c\bar{\mathbf{B}}'$ & Amplitudes & $\bar B_s^0\to\mathbf{B}_c\bar{\mathbf{B}}'$ & Amplitudes \\
\hline
$\Lambda_c^+\bar p,\,\Lambda_c^+\bar\Delta^-$ & $\mathcal{A}_{\rm ex}$ &
$\Sigma_c^{++}\bar\Delta^{--}$ & $\mathcal{A}_{\rm ex}$ \\
$\Xi_c^+\bar\Sigma^-$ & $\mathcal{A}_{\rm ex}+\mathcal{A}_{\rm em}$ &
$\Sigma_c^+\bar p,\,\Sigma_c^+\bar\Delta^-$ & $\mathcal{A}_{\rm ex}$ \\
$\Xi_c^0\bar\Lambda,\,\Xi_c^0\bar\Sigma^0$ & $\mathcal{A}_{\rm ex}+\mathcal{A}_{\rm em}$  &
$\Sigma_c^0\bar n,\,\Sigma_c^0\bar\Delta^0$ &  $\mathcal{A}_{\rm ex}$ \\
&   & $\Xi_c^{'+}\bar\Sigma^-$ &  $\mathcal{A}_{\rm ex}+\mathcal{A}_{\rm em}$ \\
&   & $\Xi_c^{'0}\bar\Lambda,\,\Xi_c^{'0}\bar\Sigma^0$  & $\mathcal{A}_{\rm ex}+\mathcal{A}_{\rm em}$ \\
&   &  $\Omega_c^0\bar\Xi^0$ & $\mathcal{A}_{\rm ex}+\mathcal{A}_{\rm em}$  \\
\hline
\end{tabular}
\end{table}
\vspace*{0.3cm}

The relevant part of the Hamiltonian for
the $\bar B^0_{(s)}\to{\bf B}_c \bar {\bf B}'$ decays
has the following form~\cite{Buras:1998raa}:
\begin{eqnarray}
\mathcal{H}_{\rm eff} =\frac{G_F}{\sqrt{2}}
\sum_{q=d,s}V_{cb}V^*_{uq}\left[c_1^{\rm eff}(\bar qu)(\bar cb)+c_2^{\rm eff}(\bar cu)(\bar qb)\right]\,,
\label{eq:EffH}
\end{eqnarray}
where $G_F$ is the Fermi constant, $V_{ij}$ stand for the CKM matrix elements, and
$(\bar q_1q_2)_{V-A}\equiv \bar q_1\gamma^\mu(1-\gamma_5)q_2$.
In the factorization approach,
the $W$-exchange amplitude of $\bar B_{(s)}^0\to{\bf B}_c\bar{\bf B}'$ is
given by~\cite{Cheng:2001ub,Hsiao:2014zza}
\begin{eqnarray}\label{amp_ex}
{\cal A}(\bar B_{(s)}^0\to{\bf B}_c\bar{\bf B}')
&=& \frac{G_F}{\sqrt{2}} \,a_2V_{cb}V^*_{uq}
\langle{\bf B}_c\bar{\bf B}'|(\bar cu)_{V-A}|0\rangle
\langle 0|(\bar qb)_{V-A}|\bar B_{(s)}^0\rangle\,,
\end{eqnarray}
where $q=d(s)$ for $\bar B^0_{(s)}$,
and $a_2=c^{\rm eff}_2+c^{\rm eff}_1/N_c$
consists of the effective Wilson coefficients
$(c_1^{\rm eff},c_2^{\rm eff})=(1.168,-0.365)$ and
the color number $N_c$.

The matrix elements in Eq.~(\ref{amp_ex}) are defined as~\cite{pdg,Cheng:2002sa}
%
\begin{eqnarray}\label{ffs}
\langle 0|\bar q\gamma^\mu(1-\gamma_5)b|B\rangle&=&-if_B q^\mu\,,\nonumber\\
\langle\mathbf{B}_c\bar{\mathbf{B}}'|\bar c\gamma^\mu u|0\rangle&=&
\bar u\left[f_1\gamma^\mu+\frac{if_2}{m_{{\bf B}_c}+m_{\bar{\bf B}'}}\sigma^{\mu\nu}q_\nu
+\frac{f_3}{m_{{\bf B}_c}+m_{\bar{\bf B}'}}q^\mu\right]v\,,\nonumber\\
\langle\mathbf{B}_c\bar{\mathbf{B}}'|\bar c\gamma^\mu\gamma_5 u|0\rangle&=&
\bar u\left[g_1\gamma^\mu+\frac{ig_2}{m_{{\bf B}_c}+m_{\bar{\bf B}'}}\sigma^{\mu\nu}q_\nu
+\frac{g_3}{m_{{\bf B}_c}+m_{\bar{\bf B}'}}q^\mu\right]\gamma_5 v\,,
\end{eqnarray}
where $f_{B}$ is the $B$ meson decay constant, 
$q^\mu=(p_{{\bf B}_c}+p_{\bar {\bf B}'})^\mu$ the momentum transfer,
and $f_i$ and $g_i$~($i=1,\,2,\,3$) are
the timelike baryonic ($0\to{\bf B}_c\bar {\bf B}'$) form factors.
The decay amplitude of $B\to{\bf B}_c\bar {\bf B}'$
in the general form is written as
\begin{eqnarray}
{\cal A}(B\to{\bf B}_c\bar{\bf B}')&=&
\bar u(A_S+A_P\gamma_5)v\,,
\label{eq:Amplitude}
\end{eqnarray}
with $A_{S,P}$ standing for the $(S,P)$-wave amplitudes.
%
By only receiving the $W$-exchange contributions in Eq.~(\ref{amp_ex}),
the $A_{S,P}$ are given by
\begin{equation}
A_S=C_S m_-\bigg[f_1+\left(\frac{m_B^2}{m_+ m_-}\right)f_3\bigg]\,,\,\,\,\,\,\,\,\,\,
A_P= C_P m_+\bigg[g_1+\left(\frac{m_B^2}{m_+^2}\right)g_3\bigg]\,,
\label{eq:SandP}
\end{equation}
where
$C_{S,P}=\mp i{G_F}\,a_2V_{cb}V^*_{uq} f_B/{\sqrt{2}}$,
$m_\pm=m_{{\bf B}_c}\pm m_{\bar {\bf B}'}$, and
$(f_2,g_2)$ vanish due to the contraction of $\sigma_{\mu\nu}q^\mu q^\nu=0$.
We use the timelike baryonic form factors
with the light-front quark model~\cite{Lih_Lcpbar},
which have been widely applied to
the $b$ and $c$-hadron decays~\cite{Bakker:2003up,
Ji:2000rd,Bakker:2002aw,Choi:2013ira,Cheng:2004cc,
Cheng:2003sm,Schlumpf:1992vq}.
In the light-front frame, since
the momentum transfer $q^\mu$ is presented as
$q^+=q^0+q^3=0$, it simply indicates
the non-contributions from $(f_3,g_3)$
as in the case of the $B\to M$ transition~\cite{Cheng:2003sm,Schlumpf:1992vq}.
%
This is in accordance with lattice QCD calculations,
where the derivations of $(f_3,g_3)\propto (f_1,g_1)/t$
show the suppression in the spacelike region~\cite{Meinel:2016dqj},
even though the corrections beyond leading order have been considered.
Therefore, $(f_3,g_3)$ are often neglected
in the literature~\cite{Cheng:2002sa,Khodjamirian:2011jp}.

By the double-pole parameterization,
$F(t)\equiv (f_1(t),g_1(t))$ is presented as~\cite{Khodjamirian:2011jp}
\begin{eqnarray}\label{q_square}
F(t)=\frac{F(0)}{1-a\left(t/m_B^2\right)+b\left(t^2/m_B^4\right)}\,,
\end{eqnarray}
with $t\equiv q^2$.
%
The $SU(3)$ flavor symmetry can relate $F(0)$
in the different decays, given by
\begin{eqnarray}\label{eq:FF}
(C_f^{{\bf B}_c\bar{\bf B}'}, C_g^{{\bf B}_c\bar{\bf B}'})
=\xi (C_f^{\Lambda^+_c\bar p},C_g^{\Lambda^+_c\bar p})\,,\;
(C_f^{{\bf B}_c\bar{\bf B}'}, C_g^{{\bf B}_c\bar{\bf B}'})
=\zeta (C_f^{\Sigma^+_c\bar p},C_g^{\Sigma^+_c\bar p})\,,
\end{eqnarray}
with $(C_f, C_g)\equiv (f_1(0),g_1(0))$ and
$|\xi|$ ($|\zeta|$) for the anti-triplet (sextet) charmed baryons
listed in Table~\ref{tab:ffrelations}.
%
\begin{table}[t!]
\caption{The relations between different timelike baryon form factors.
The $\bar{\mathbf{3}}_c$ and $\mathbf{6}_c$ denote
the anti-triplet and the sextet charmed baryons, respectively,
and the $\mathbf{8}$ denotes the octet charmless baryons.}\label{tab:ffrelations}
\begin{center}
\begin{tabular}{lrclr}
\hline
      $\bar{\mathbf{3}}_c\otimes\mathbf{8}$                            & $|\xi|$                        &~~~~~~& $\mathbf{6}_c\otimes\mathbf{8}$     &   $|\zeta|$    \\
\hline
$\Lambda^+_c\bar p$      &  $1$  &  &  $\Sigma_c^{+}\bar p$       &  1    \\
$\Xi_c^+\bar\Sigma^-$ &  $1$  &  &  $\Sigma_c^{0}\bar n$       &  $\sqrt{2}$                 \\
$\Xi_c^0\bar\Lambda$ &  $\frac{1}{\sqrt{6}}$             &  &  $\Xi_c^{'+}\bar\Sigma^-$  &  1   \\
$\Xi_c^0\bar\Sigma^0$ & $\frac{1}{\sqrt{2}}$    &  &  $\Xi_c^{'0}\bar\Sigma^0$ &  $\frac{1}{\sqrt{2}}$            \\
                                      &                                  &  &  $\Xi_c^{'0}\bar\Lambda$  &  $\sqrt{\frac{3}{2}}$   \\
                                      &                                  &  &  $\Omega_c^0\bar\Xi^0$   &  $\sqrt{2}$               \\
\hline
\end{tabular}
\end{center}
\end{table}
%
%
We compute the branching ratios from the decay-rate equation for two-body decays, given by~\cite{pdg}
\begin{eqnarray}\label{p_space}
&&{\cal B}(B\to{\bf B}_c\bar {\bf B}')=
\frac{|\vec{p}_{{\bf B}_c}|\tau_{B}}{8\pi m_{B}^2 }|{\cal A}(B\to{\bf B}_c\bar {\bf B}')|^2\,,\nonumber\\
&&|\vec{p}_{{\bf B}_c}|=\frac{
\sqrt{(m_{B}^2-m_+^2)(m_{B}^2-m_-^2)}}{2 m_{B}}\,,
\end{eqnarray}
where $\tau_{B}$ denotes the $B$ meson lifetime.

%
\section{Numerical analysis}
In the numerical analysis, we adopt the Wolfenstein parameterization
for the CKM matrix elements, given by~\cite{pdg}
\begin{eqnarray}
(V_{cb},V_{ud},V_{us})=(A\lambda^2,1-\lambda^2/2,\lambda)\,,
\end{eqnarray}
where $\lambda=0.22453\pm 0.00044$ and $A=0.836\pm 0.015$,
together with the $B$ meson decay constants
$(f_B,f_{B_s})=(0.19,0.23)$~GeV~\cite{pdg}.
%
The light-front quark model provides
\begin{eqnarray}\label{q_square2}
&&(C_f^{\Lambda^+_c\bar p},a,b)=(0.41^{+0.07}_{-0.09},0.75,0.45)\,,\nonumber\\
&&(C_g^{\Lambda^+_c\bar p},a,b)=(0.37^{+0.07}_{-0.08},0.90,0.50)\,,\nonumber\\
&&(C_f^{\Sigma^+_c\bar p},a,b)=(-0.25\pm0.04,0.70,0.40)\,,\nonumber\\
&&(C_g^{\Sigma^+_c\bar p},a,b)=(-0.21\pm0.04,0.84,0.52)\,,
\end{eqnarray}
where $C_{f,g}^{\Lambda^+_c\bar p}$ and $C_{f,g}^{\Sigma^+_c\bar p}$
relate the different decays with $(|\xi|,|\zeta|)$ in Table~\ref{tab:ffrelations}.

By considering the non-factorizable $W$-emission contributions only,
the pole model relates
$B^-\to \Sigma_c^0 \bar p$ and $\bar B^0\to \Lambda_c^+ \bar p$
with the same sets of strong coupling constants~\cite{Cheng:2002sa,Bevan:2014iga}.
We follow Ref.~\cite{Cheng:2002sa} to re-extract the strong coupling constants
with the most current data of
${\cal B}(B^-\to \Sigma_c^0 \bar p)=2.9\times 10^{-5}$~\cite{pdg},
which leads to ${\cal B}(\bar B^0\to \Lambda_c^+ \bar p)=0.4\times 10^{-5}$
in comparison with the data of $(1.55\pm 0.18)\times 10^{-5}$.
In order to explain the data,
the often neglected $W$-exchange contribution to ${\cal B}(\bar B^0\to \Lambda_c^+ \bar p)$
should be at the level of $10^{-5}$.
With $N_c=2.9$ ($a_2=0.04$), we are able to get
${\cal B}(\bar B^0\to \Lambda_c^+ \bar p)=(1.0^{+0.4}_{-0.3})\times 10^{-5}$
with the errors from the uncertainties of the form factors.
In the generalized factorization,
one allows $N_c$ to shift from 2 to $\infty$ as an effective number
to account for the non-factorizable strong interaction. Particularly,
$N_c\simeq 3$ implies a mild correction~\cite{ali}.
Accordingly, we estimate the previously neglected $W$-exchange contributions
to ${\cal B}(B\to{\bf B}_c \bar {\bf B}')$,
given in Table~\ref{tab:BRs}.
%
\begin{table}[t!]
\caption{Branching ratios of the $\bar B^0_{(s)}\to{\bf B}_c\bar {\bf B}'$ decays
from the $W$-exchange diagrams.} \label{tab:BRs}
\begin{center}
\begin{tabular}{lcccclccc}
\hline
Decay modes& $\mathcal{B}\times10^{5}$ &~~~&  Data &   &
Decay modes&   $\mathcal{B}\times10^{6}$ &~~~& Data  \\
\hline
$\bar B^0\to\Lambda^+_c\bar p$ &  $1.0^{+0.4}_{-0.3}$  &  & $1.54\pm0.18$~\cite{pdg}  &   &
$\bar B^0\to\Sigma_c^{+}\bar p$   & $2.9^{+0.8}_{-0.9}$  & & $<24$~\cite{pdg} \\
$\bar B^0\to\Xi_c^+\bar\Sigma^-$ & $1.1\pm 0.4$ & & & &
$\bar B^0\to\Sigma_c^{0}\bar n$   &  $5.8^{+1.5}_{-1.8}$ & & \\
$\bar B^0\to\Xi_c^0\bar\Sigma^0$ & $0.6\pm 0.2$ & & & &
$\bar B^0\to\Xi_c^{'+}\bar\Sigma^-$ & $3.1^{+0.9}_{-1.0}$ & & \\
$\bar B^0\to\Xi_c^0\bar\Lambda$ &  $0.2\pm0.1$ & & & &
$\bar B^0\to\Xi_c^{'0}\bar\Sigma^0$ & $1.6^{+0.4}_{-0.5}$ & & \\&   &  & & &
$\bar B^0\to\Xi_c^{'0}\bar\Lambda$ & $4.6^{+1.3}_{-1.5}$ & & \\&   &  & &  &
$\bar B^0\to\Omega_c^0\bar\Xi^0$ &  $6.4^{+1.8}_{-2.1}$ & & \\
\hline%
\hline
Decay modes& $\mathcal{B}\times10^{6}$ &~~~&     &   &
Decay modes&   $\mathcal{B}\times10^{7}$ &~~~&   \\
\hline
$\bar B_s^0\to\Lambda^+_c\bar p$ &  $0.8\pm 0.3$  &  &  &   &
    $\bar B_s^0\to\Sigma_c^{+}\bar p$   & $2.3^{+0.6}_{-0.7}$  & &\\
$\bar B_s^0\to\Xi_c^+\bar\Sigma^-$ & $0.9\pm 0.3$ & & & &
    $\bar B_s^0\to\Sigma_c^{0}\bar n$   &  $4.5^{+1.1}_{-1.4}$ & &\\
$\bar B_s^0\to\Xi_c^0\bar\Sigma^0$ & $0.4\pm 0.2$ & & & &
    $\bar B_s^0\to\Xi_c^{'+}\bar\Sigma^-$ & $2.5^{+0.6}_{-0.8}$ & &\\
$\bar B_s^0\to\Xi_c^0\bar\Lambda$ &  $0.2\pm0.1$ & & & &
    $\bar B_s^0\to\Xi_c^{'0}\bar\Sigma^0$ & $1.2^{+0.3}_{-0.4}$ & &\\& &  &  & &
    $\bar B_s^0\to\Xi_c^{'0}\bar\Lambda$ & $3.6^{+0.9}_{-1.2}$ & &\\& &  &  & &
    $\bar B_s^0\to\Omega_c^0\bar\Xi^0$ &  $5.1^{+1.4}_{-1.7}$ & &\\
\hline
\end{tabular}
\end{center}
\end{table}
%

\newpage
\section{Discussions and Conclusions}
Several suppression factors have been proposed
to support the neglecting of the $W$-exchange (annihilation) contributions.
The first one
is in analogy with the semileptonic $B\to \ell\bar \nu_\ell$ decays~\cite{Bevan:2014iga}.
In fact, we obtain
\begin{eqnarray}\label{helicity_br}
{\cal B}(B\to \ell\bar \nu_\ell)&=&
\frac{G_F^2 m_B \tau_B}{8\pi}|V_{ub}|^2 f_B^2 m_\ell^2
\bigg(1-\frac{m_\ell^2}{m_B^2}\bigg)^2\,,\nonumber\\
{\cal B}(B\to {\bf B}_c\bar {\bf B}')&=&
\frac{G_F^2 |\vec{p}_{{\bf B}_c}|\tau_B}{8\pi}
|a_2V_{cb}V_{uq}^*|^2 f_{B}^2 m_+^2
 \bigg[ {\cal R}_m f_1^2 \bigg(1-\frac{m_+^2}{m_B^2}\bigg)
+ g_1^2 \bigg(1-\frac{m_-^2}{m_B^2}\bigg)\bigg]\,,
\end{eqnarray}
with ${\cal R}_m\equiv (m_-/m_+)^2$.
While $m_\mu^2$ is responsible for the helicity suppressed
${\cal B}(B^-\to \mu^-\bar \nu_\mu)\simeq 10^{-7}$~\cite{Hou:2019uxa},
$B\to {\bf B}_c\bar {\bf B}'$ with
$m_+^2=(m_{{\bf B}_c}+m_{\bar {\bf B}'})^2$ obviously allows for helicity-flip.
The second one is
the decay constant $f_B=0.19$~GeV in Eq.~(\ref{helicity_br}),
regarded as small and suppressing the $W$-exchange contribution~\cite{Jarfi:1990ej}.
However, it is found that
${\cal B}(B^-\to\tau^-\bar \nu_\tau)$ with $f_B$ can still be as large as
$(1.09\pm 0.24)\times 10^{-4}$~\cite{pdg}.
Besides, the ratio of $m_{\mu}^2/m_{\tau}^2\sim {\cal O}(10^{-3})$
implies the smallness of
${\cal B}(B^-\to\mu\bar \nu_\mu)/{\cal B}(B^-\to\tau\bar \nu_\tau)\sim {\cal O}(10^{-3})$.
Hence, it is reasonable to infer that
$f_B$ cannot be the main cause of the suppression.
Third,
in the {\it charmless} $B\to{\bf B}\bar {\bf B}'$ decays,
one of the suppression factors
for ${\cal B}(\bar B^0\to p\bar p)\sim {\cal O}(10^{-8})$ 
comes from $(f_1,g_1)\propto (\alpha_s/t)^2$ with $t=m_B^2$,
which corresponds to the hard gluons
that transfer the energy of $m_B$~\cite{Hsiao:2014zza,Hsiao:2018umx}.
In the $B\to{\bf B}_c \bar {\bf B}'$ decays, however,
$(f_1,g_1)$ with a charmed baryon are not small at $t=m_B^2$,
as shown in Eqs.~(\ref{q_square}) and (\ref{q_square2}).
Note that $(f_1,g_1)$ in the timelike and spacelike regions can be associated
with the crossing symmetry, and $(f_1,g_1)$ with the light-front quark model
agree with those in lattice QCD calculations~\cite{Lih_Lcpbar,Meinel:2016dqj}.
By contrast,
${\cal B}(\bar B^0\to\Lambda_c^+\bar p)$ was once calculated
as small as $4.6\times 10^{-7}$~\cite{Korner:1988mx},
taken as the another theoretical support
for the neglect of the $W$-exchange diagram~\cite{Cheng:2002sa,Bevan:2014iga}.
However, the estimation was done with
the baryonic form factors in the dipole form of
$F(0)/(1-t/m_{D^*}^2)^2$ adopted from the $\Lambda_c^+\to p$ transition~\cite{Korner:1988mx},
such that the momentum transfer at $m_B$ 
exceeds the $D^*$ meson pole, causing the suppression,
whereas the validity of the $D^*$ meson pole has never been tested
in the timelike region.
Besides, $F(0)$ was not clearly given,
due to the lack of the studies on the quark and QCD models at that time.

With the $W$-exchange contributions,
we obtain
\begin{eqnarray}\label{cal_Lcpbar}
{\cal B}(\bar B^0\to\Sigma_c^{+}\bar p)&=&(2.9^{+0.8}_{-0.9})\times 10^{-6}\,,\nonumber\\
{\cal B}(\bar B^0\to\Sigma_c^{0}\bar n)&=&(5.8^{+1.5}_{-1.8})\times 10^{-6}\,,
\end{eqnarray}
where ${\cal B}(\bar B^0\to\Sigma_c^{+}\bar p)$ is
consistent with the current data in Eq.~(\ref{BcBdata}).
In particular, ${\cal B}(\bar B^0\to\Sigma_c^{0}\bar n)$
is predicted to be ten times larger than the pole model calculation~\cite{Cheng:2002sa}.
On the other hand, by only considering the $W$-emission contribution,
the pQCD approach gives that
${\cal B}(\bar B^0\to \Lambda_c^+ \bar p)=(2.3-5.1)\times 10^{-5}$~\cite{He:2006vz},
which is more than 4 standard deviations away
from the measured central value.
It is interesting to know if the pQCD approach would
overestimate ${\cal B}(B^-\to \Sigma_c^0 \bar p)$ as well.
%
One way of comparing the $W$-emission and exchange contributions
is by studying decays other than $\bar B^0\to \Lambda_c^+\bar p,\Sigma_c^+\bar p$,
which receive contributions from both ${\cal A}_{\rm{ex}}$ and ${\cal A}_{em}$.
By taking ${\cal A}_{\rm{ex}}$ as the primary contribution,
we predict that
\begin{eqnarray}
&&{\cal B}(\bar B^0\to\Xi_c^0\bar\Sigma^0,\Xi_c^0\bar\Lambda)
=(0.6\pm 0.2,0.2\pm 0.1)\times 10^{-5}\,,\nonumber\\
&&{\cal B}(\bar B^0_s\to\Xi_c^0\bar\Sigma^0,\Xi_c^0\bar\Lambda)
={\cal R}_{B_s}\times{\cal B}(\bar B^0\to\Xi_c^0\bar\Sigma^0,\Xi_c^0\bar\Lambda^0)\,,
\end{eqnarray}
with ${\cal R}_{B_s}\equiv |(V_{us}f_{B_s})/(V_{ud}f_{B})|^2=0.075$.
In future measurements
the deviation of ${\cal R}_{B_s}$ from the theoretical prediction
can be used to evaluate the size of ${\cal A}_{em}$ in the decays.
Being pure $W$-exchange processes,
$\bar B^0\to\Xi_c^+\bar \Sigma^-,\Xi_c^{'+}\bar \Sigma^-,
\Omega_c^0\bar\Xi^0$ and
$\bar B^0_s\to\Sigma_c^0\bar n, \Lambda_c^+\bar p,
\Sigma_c^+\bar p$ can be excellent probes for the examination of
the $W$-exchange mechanism.
Therefore, we predict that
\begin{eqnarray}
{\cal B}(\bar B^0\to\Xi_c^+\bar\Sigma^-)&=&(1.1\pm 0.4)\times 10^{-5}\,,\nonumber\\
{\cal B}(\bar B_s^0\to\Lambda^+_c\bar p)&=&(0.8\pm 0.3)\times 10^{-6}\,, \nonumber\\
{\cal B}(\bar B^0\to \Xi_c^{'+}\bar\Sigma^-,\Omega_c^0\bar\Xi^0)
&=& (3.1^{+0.9}_{-1.0},6.4^{+1.8}_{-2.1})\times10^{-6}\,,
\end{eqnarray}
which are all within the capability of the current $B$ factories.

In summary, we have studied the charmful two-body baryonic
$B_{(s)}^0\to {\bf B}_c \bar {\bf B}'$ decays,
with ${\bf B}_{c}=(\Xi_c^{+,0},\Lambda_c^+)$ or
$(\Sigma_c^{++,+,0},\Xi_c^{\prime +,0},\Omega_c^0)$.
We have found that the often neglected
$W$-exchange contribution is in fact not negligible.
%
Particularly, we obtain that
${\cal B}(\bar B^0\to\Sigma_c^{+}\bar p)=(2.9^{+0.8}_{-0.9})\times 10^{-6}$,
agreeing with the current data.
For decays that only proceed via the $W$-exchange diagram,
we predict
${\cal B}(\bar B_s^0\to\Lambda^+_c\bar p)=(0.8\pm 0.3)\times 10^{-6}$ and
${\cal B}(\bar B^0\to\Xi_c^+\bar\Sigma^-)=(1.1\pm 0.4)\times 10^{-5}$,
which provide an excellent window for the LHCb and Belle II experiments to look into
the long-ignored $W$-exchange mechanism.

\section*{ACKNOWLEDGMENTS}
This work was supported in part by
National Science Foundation of China (11675030) and
U. S. National Science Foundation award ACI-1450319.


\begin{thebibliography}{99}

\bibitem{pdg}
M.~Tanabashi {\it et al.} [Particle Data Group], 
Phys.\ Rev.\ D {\bf 98}, 030001 (2018).

\bibitem{Wei:2007fg}
J.T.~Wei {\it et al.} [Belle Collaboration],
Phys.\ Lett.\ B {\bf 659}, 80 (2008). 

\bibitem{Aaij:2017vnw}
R.~Aaij {\it et al.} [LHCb Collaboration],
Phys.\ Rev.\ Lett.\  {\bf 119}, 041802 (2017). 

\bibitem{Aaij:2017pgn}
R.~Aaij {\it et al.} [LHCb Collaboration],
Phys.\ Rev.\ D {\bf 96}, 051103 (2017). 

\bibitem{Lu:2018qbw}
P.C.~Lu {\it et al.} [Belle Collaboration],
Phys.\ Rev.\ D {\bf 99}, 032003 (2019). 

\bibitem{Hou:2000bz}
W.S.~Hou and A.~Soni, Phys.\ Rev.\ Lett.\  {\bf 86}, 4247 (2001).

\bibitem{Suzuki:2006nn}
M.~Suzuki,
J.\ Phys.\ G {\bf 34}, 283 (2007). 

\bibitem{Aaij:2017gum}
R.~Aaij {\it et al.} [LHCb Collaboration],
Phys.\ Rev.\ Lett.\  {\bf 119}, 232001 (2017). 

\bibitem{Aaij:2016xfa}
R.~Aaij {\it et al.} [LHCb Collaboration],
JHEP {\bf 04}, 162 (2017). 

\bibitem{Aaij:2013fla}
R.~Aaij {\it et al.} [LHCb Collaboration],
Phys.\ Rev.\ D {\bf 88}, 052015 (2013). 










\bibitem{Chen:2008pf}
C.H.~Chen, H.Y.~Cheng and Y.K.~Hsiao,
Phys.\ Lett.\ B {\bf 663}, 326 (2008). 

\bibitem{Bevan:2014iga}
A.J.~Bevan {\it et al.} [BaBar and Belle Collaborations],
Eur.\ Phys.\ J.\ C {\bf 74}, 3026 (2014).

\bibitem{Chernyak:1990ag}
V.~L.~Chernyak and I.~R.~Zhitnitsky,
Nucl.\ Phys.\ B {\bf 345}, 137 (1990).

\bibitem{Ball:1990fw}
P.~Ball and H.~G.~Dosch,
Z.\ Phys.\ C {\bf 51}, 445 (1991).

\bibitem{Cheng:2001tr}
H.Y.~Cheng and K.C.~Yang,
Phys.\ Rev.\ D {\bf 66}, 014020 (2002). 


\bibitem{Chang:2001jt}
C.H.V.~Chang and W.S.~Hou,
Eur.\ Phys.\ J.\ C {\bf 23}, 691 (2002). 

\bibitem{Chua:2013zga}
C.K.~Chua, 
Phys.\ Rev.\ D {\bf 89}, 056003 (2014).


\bibitem{He:2006vz}
X.~G.~He, T.~Li, X.~Q.~Li and Y.~M.~Wang,
Phys.\ Rev.\ D {\bf 75}, 034011 (2007)




\bibitem{Gabyshev:2002dt}
N.~Gabyshev {\it et al.} [Belle Collaboration],
Phys.\ Rev.\ Lett.\  {\bf 90}, 121802 (2003). 

\bibitem{Aubert:2008ax}
B.~Aubert {\it et al.} [BaBar Collaboration],
Phys.\ Rev.\ D {\bf 78}, 112003 (2008).




\bibitem{Cheng:2001ub}
H.Y.~Cheng and K.C.~Yang,
Phys.\ Rev.\ D {\bf 65}, 054028 (2002)
Erratum: [Phys.\ Rev.\ D {\bf 65}, 099901 (2002)].


\bibitem{Cheng:2002sa}
H.Y.~Cheng and K.C.~Yang,
Phys.\ Rev.\ D {\bf 67}, 034008 (2003). 



\bibitem{Hsiao:2014zza}
Y.K.~Hsiao and C.Q.~Geng,
Phys.\ Rev.\ D {\bf 91}, 077501 (2015). 


\bibitem{Bediaga:1991eu}
I.~Bediaga and E.~Predazzi, 
Phys.\ Lett.\ B {\bf 275}, 161 (1992).

\bibitem{Pham:1980dc}
X.Y.~Pham,
Phys.\ Rev.\ Lett.\  {\bf 45}, 1663 (1980).

\bibitem{Pham:1980xe}
X.Y.~Pham, 
Phys.\ Lett.\  {\bf 94B}, 231 (1980).

\bibitem{Buras:1998raa} 
A.J.~Buras, hep-ph/9806471.

\bibitem{Lih_Lcpbar}
Chong-Chung Lih and Y.K. Hsiao,
``Baryonic form factors involving a charmed baryon in the light-front quark model,''
in preparation.




\bibitem{Bakker:2003up}
B.L.G.~Bakker, H.M.~Choi and C.R.~Ji,
Phys.\ Rev.\ D {\bf 67}, 113007 (2003). 

\bibitem{Ji:2000rd}
C.R.~Ji and C.~Mitchell,
Phys.\ Rev.\ D {\bf 62}, 085020 (2000).

\bibitem{Bakker:2002aw}
B.L.~G.~Bakker and C.R.~Ji,
Phys.\ Rev.\ D {\bf 65}, 073002 (2002). 

\bibitem{Choi:2013ira}
H.M.~Choi and C.R.~Ji,
Few Body Syst.\  {\bf 55}, 435 (2014). 

\bibitem{Cheng:2004cc}
H.Y.~Cheng, C.K.~Chua and C.W.~Hwang,
Phys.\ Rev.\ D {\bf 70}, 034007 (2004). 

\bibitem{Cheng:2003sm}
H.Y.~Cheng, C.K.~Chua and C.W.~Hwang,
Phys.\ Rev.\ D {\bf 69}, 074025 (2004). 

\bibitem{Schlumpf:1992vq}
F.~Schlumpf,
Phys.\ Rev.\ D {\bf 47}, 4114 (1993)
Erratum: [Phys.\ Rev.\ D {\bf 49}, 6246 (1994)]. 

\bibitem{Meinel:2016dqj}
S.~Meinel,
Phys.\ Rev.\ Lett.\  {\bf 118}, 082001 (2017).

\bibitem{Khodjamirian:2011jp}
A.~Khodjamirian, C.~Klein, T.~Mannel and Y.M.~Wang,
JHEP {\bf 09}, 106 (2011). 

\bibitem{ali} A. Ali, G. Kramer, and C.D. Lu, Phys. Rev.  ${\bf D 58}$, 094009 (1998).

\bibitem{Hou:2019uxa}
W.S.~Hou, M.~Kohda, T.~Modak and G.G.~Wong,
Phys.\ Lett.\ B {\bf 800}, 135105 (2020). 

\bibitem{Jarfi:1990ej}
M.~Jarfi, O.~Lazrak, A.~Le Yaouanc, L.~Oliver, O.~Pene and J.C.~Raynal,
Phys.\ Rev.\ D {\bf 43}, 1599 (1991).

\bibitem{Hsiao:2018umx}
Y.K.~Hsiao, C.Q.~Geng, Y.~Yu and H.J.~Zhao,
Eur.\ Phys.\ J.\ C {\bf 79}, 433 (2019). 

\bibitem{Korner:1988mx}
J.G.~Korner,
Z.\ Phys.\ C {\bf 43}, 165 (1989).














\end{thebibliography}
\end{document}